\documentclass{jfm}
\usepackage{graphicx}
\usepackage{epstopdf, epsfig}
\usepackage{color}
\usepackage{amssymb}
\usepackage{amsmath}
\usepackage{textcomp} 
\usepackage{hyperref}
\hypersetup{
colorlinks,
citecolor=blue,
linkcolor=blue}

\newcommand{\um}{\langle \mathbf u \rangle}
\newcommand{\uzm}{\langle u_z \rangle}
\newcommand{\ozm}{\langle \omega_z \rangle}
\newcommand{\pmean}{\langle p \rangle}

\shorttitle{Mean flow anisotropy without waves in rotating turbulence}
\shortauthor{J.A. Brons, P.J. Thomas and  A. Poth\'erat}

\title{Mean flow anisotropy without waves in rotating turbulence}

\author{Jonathan A. Brons\aff{1,2}
  \corresp{\email{jonbrons@gmail.com}},
  P. J.  Thomas\aff{2}
 \and A. Poth\'erat\aff{1}}

\affiliation{\aff{1}Fluid and Complex Systems Research Centre, Coventry University, Coventry CV1 5FB, UK
\aff{2} Fluid Dynamics Research Centre, School of Engineering, University of Warwick,
Coventry CV4 7AL, UK}

\begin{document}

\maketitle

\begin{abstract}
We tackle the question of how anisotropy in flows subject to background rotation favours structures elongated along the rotation axis, especially in turbulent flows. A new, wave-free mechanism is identified that challenges the current understanding of the process. Inertial waves propagating near the rotation axis \citep{staplehurst08,yarom14_nat} are generally accepted as the most efficient mechanism to transport energy anisotropically. They have been shown to transfer energy to large anisotropic, columnar structures. Nevertheless, they cannot account for the formation of simpler steady anisotropic phenomena such as Taylor columns. Here, we experimentally show that more than one mechanism involving the Coriolis force may promote anisotropy. In particular, in the limit of fast rotation, that is at low Rossby number, anisotropy favouring the direction of rotation of the average of a turbulent flow arises neither because of inertial waves nor following the same mechanism as in steady Taylor columns, but from an interplay between the Coriolis force and average advection.
\end{abstract}


\section{Introduction}
Subjecting a flow to background rotation tends to eliminate variations of velocity along the axis of rotation. The effect, first noticed by Lord Kelvin \citep{kelvin1868}, was famously illustrated when Taylor observed that a fluid column exactly followed the motion of a coin placed at the bottom of a rotating tank \citep{taylor1922}. The question of the anisotropic mechanism underlying the development of these columnar structures is, however, still open. It raises the wider issue of how anisotropy favouring the direction of rotation arises in rotating flows, which is the focus of this work. The thesis we advocate is that non-propagative mechanisms can promote anisotropy, in particular purely advective ones.\\
The type of anisotropy that underpins the emergence of Taylor columns in rotating flows is most commonly studied in the context of turbulence in fast rotating systems such as planetary cores, atmospheres and astrophysical objects, where its origin is attributed to the propagation of inertial waves (\cite{davidsonA,hopfinger1982_jfm}, and see \citet{greenspan} for the theory of these waves): for a wavevector $\mathbf k$ of frequency $\omega$, in background rotation $\mathbf \Omega$, inertial waves follow the dispersion relation, and associated group velocity
\begin{equation}
\omega=\pm2\mathbf \Omega\cdot \mathbf e_k, \qquad \mathbf v_g=\pm\frac2k \mathbf e_k\times (\mathbf \Omega\times \mathbf e_k),
\label{eq:iw_disp}
\end{equation}
where $\mathbf e_k=\frac1k\mathbf k$. Two main theories account for the growth of anisotropy favouring the direction of rotation and the subsequent spontaneous formation of large structures in these systems: 
one invokes triadic interactions feeding an inverse energy cascade towards large scales while non-resonant triads or quartets of waves transfer energy to modes aligned with the axis of rotation \citep{cambon99_anrev,smith99_pof}. This scenario is supported by numerical simulations and by strong experimental \citep{campagne14_pof,lamriben11_prl,duran13_pre} and numerical \citep{chen05_jfm,smith96_prl} evidence of the inverse energy cascade.
The other theory argues that linear inertial waves account for most of the energy transport in rotating turbulence \citep{davidson06}. This was demonstrated numerically and experimentally in the context of the propagation of transient rotating turbulence \citep{kolvin09,staplehurst08,screen08_pof}.\\
Both theories consider Taylor Columns as large turbulent scales, that is, as transient, fluctuating objects. None of them, however, satisfactorily explains the steady, laminar columns that Taylor observed. Indeed, the analytical solution for these columns \citep{moore69} is entirely steady and neglects non-rotating inertia, thus excluding inertial waves. The formation of fluctuating Taylor columns in rotating turbulent flows and steady ones in laminar flows may thus arise out of fundamentally different mechanisms. This idea finds support in previous experimental and theoretical work \citep{mcewan1973_gfd,mcewan1976_nat,p12_epl}, showing that a velocity field with a divergence in planes normal to the rotation preferentially transports momentum along the rotation axis. Crucially, this phenomenology holds regardless of the steady or fluctuating nature of the flow. We therefore suggest that more than a single mechanism may exist to promote anisotropy in rotating flows and set out to determine conditions in which the better known mechanisms involving inertial waves may not be dominant. Beyond simple steady flows, we seek evidence of such alternative mechanism not involving inertial waves in the average components of turbulent flows, on the grounds that these are both steady in nature and subject to the presence of inertial waves inherent to rotating turbulence. As such, they provide the ideal battleground for mechanisms with and without waves to compete. The particular flow of interest shall be generated in an experimental device where turbulent motion results from the fast injection of fluid through holes located at the bottom of a cylindrical tank driven in rotation. This technique, pioneered by \cite{mcewan1973_gfd}, provides an efficient and convenient mean of generating turbulence with a background rotation (see more recent experiments by \cite{kolvin09}). For the specific purpose of the study, using a small number of holes enables us to imprint a strong average component to the turbulence. \\

We first establish a scaling law for the length scale of anisotropic structures forming out of momentum transport associated to average flows with a divergence in planes normal to the rotation (section \ref{sec:theory}). The experimental setup and optical measurement techniques based on Particle Image Velocimetry are detailed in section \ref{sec:experiment}. We then proceed in three steps gathered in section \ref{sec:results}: first, the theoretical scaling for the length scale of columnar structures is validated against experimental measurements. Then, in order to separate the contribution of random fluctuations from that of inertial waves to the topology of the average flow, we introduce a filtering technique using the maximum frequency of inertial waves.  Based on this technique, we derive mathematical upper bounds for terms involving inertial waves in the equation governing the average velocity field. The relevance of these upper bounds is tested by means of more refined filtering techniques inspired from \cite{yarom14_nat}. Finally we derive an equation for the average flow and experimentally estimate upper bounds for terms involving inertial waves to identify flow regimes where the contributions of these waves in this equation can be neglected. The implications of these results are discussed in  section \ref{sec:conclusion}.

\section{Theory}
\label{sec:theory}
\subsection{Governing equations and diffusion length scale}
We first derive scalings characterising anisotropy between the rotation axis and other directions in steady and turbulent flows. Consider an incompressible flow of Newtonian fluid in a frame of reference rotating at constant angular velocity $\Omega \mathbf e_z$. The effect of the Coriolis force on a structure of size $l_z$ along the axis of rotation, $l_\perp$ in the directions perpendicular to it and velocity $U$ is readily seen from the $z-$component of the vorticity equation governing the velocity and vorticity fields $\mathbf u$ and $\boldsymbol{\omega}$:
\begin{equation}
\left(\frac{d}{dt} -\nu\Delta\right)\omega_z= \boldsymbol{\omega} \cdot \nabla u_z+2\Omega\partial_z u_z,
\label{eq:omegaz}
\end{equation}
where, $d/dt=\partial_t+\mathbf u\cdot \nabla$. In the limit $\Omega\rightarrow\infty$ the flow is columnar, with $\partial_zu_z= 0$, which implies $ -\nabla_\perp\cdot\mathbf u_\perp=0$. For finite rotation, a horizontally divergent flow exists and the Coriolis force associated to it must be balanced either by inertial or viscous forces \citep{p12_epl}. The divergent flow is estimated by means of the $z$-component and the divergence of the momentum equation i.e.: 
\begin{eqnarray}
\left(\frac{d}{dt} -\nu\Delta\right)u_z&=&-\partial_z \frac{p}\rho, 
\label{eq:nsz}\\
\nabla\cdot\left(\mathbf u\cdot\nabla\mathbf u \right)&=&2\Omega\omega_z-\Delta \frac{p}\rho.
\label{eq:nsp}
\end{eqnarray}
In both Taylor's experiment \citep{taylor1922} and Moore \& Saffman's analytical solution \citep{moore69}, inertia is neglected. In this limit (\ref{eq:nsp}) implies that the pressure is geostrophic $p=2\rho\Omega\Delta^{-1}\omega_z$, where the inverse of the Laplacian $\Delta^{-1}$ is defined with boundary conditions prescribed by the geometry. The rotational part of the Coriolis force can thus be expressed by virtue of (\ref{eq:nsz}) as 
\begin{equation}
2\Omega\partial_zu_z=4\frac{\Omega^2}\nu\partial^2_{zz}\Delta^{-2}\omega_z. 
\label{eq:coriolis_viscous}
\end{equation}
An almost identical mathematical form exists for the Lorentz force in electrically conducting fluids pervaded by an imposed magnetic field $B\mathbf e_z$, where it expresses that the Lorentz force diffuses momentum along $\mathbf e_z$ \citep{sm82}. This finding was experimentally verified, establishing that the diffusive nature of the Lorentz force persists both in viscous and inertial regimes, albeit with different characteristic diffusion length scales \citep{sm82,pk2014_jfm,bpdd2018_prl}.  
In the rotating flows of interest here, the Coriolis force diffuses momentum along $\mathbf e_z$ in the inertialess limit. Its diffusion length scale follows from introducing (\ref{eq:coriolis_viscous}) into (\ref{eq:omegaz}) and applying scaling arguments:
\begin{equation}
l_z^\nu(l_\perp)\sim l_\perp\frac{2\Omega l_\perp^2}{\nu}= l_\perp\frac{l_\perp^2}{H^2} E^{-1},
\end{equation}
where the Ekman number $E=\nu/2\Omega H^2$ represents the ratio of Coriolis to viscous forces, based on the domain height $H$ (cf. figure \ref{fig1_setup}). This length scale recovers the columnar length scale implied in \citet{moore69}'s analytical solution. $l_z^\nu$ can be interpreted as the distance needed for viscous effects to exhaust the horizontally divergent flow that drives the column.\\
In contrast to Taylor's flow \citep{taylor1922}, inertia dominates in turbulent flows and balances the Coriolis force associated to the horizontally divergent flow in (\ref{eq:omegaz}). Using this assumption and a similar derivation as for $l_z^\nu$ leads to an inertial scaling for $l_z$:
\begin{equation}
l_z^I(l_\perp)\sim\frac{2\Omega l_\perp^2}{U}= l_\perp Ro(l_\perp)^{-1},
\label{eq:lz_inertia}
\end{equation}
where the Rossby number $Ro(l_\perp)=U/2\Omega l_\perp$ is based on the considered structure's velocity scale $U$. It represents the ratio of inertial to Coriolis forces at the scale of the structure considered.\\

\subsection{Influence from the fluctuating flow on the average flow component}
To isolate the mechanisms controlling anisotropy, we consider a forced, anisotropic turbulent flow with non-zero average flow at large Reynolds number. A benefit of this choice of flow is that mechanisms controlling the anisotropy of the average flow that do not involve waves, as in Taylor columns, can be captured by simple event-averaging. At the same time, since turbulent fluctuations under strong rotation support inertial waves, these can potentially affect the anisotropy of the average flow. For these reasons, a turbulent flow with an average flow component offers a good testing ground to identify the conditions in which either propagative or wave-free mechanisms drive anisotropy. We start by deriving the equations for the average quantities, decomposing all quantities into their average and fluctuations, \emph{e.g.} $\mathbf u=\um +\mathbf u^\prime$. Taking the average of (\ref{eq:omegaz})-(\ref{eq:nsp}), neglecting viscous friction yields:
\begin{equation}
\um \cdot\nabla \ozm= \langle\boldsymbol{\omega}\rangle\cdot\nabla \uzm+ 2\Omega\partial_z \uzm+ \langle\boldsymbol{\omega}^\prime\cdot\nabla u_z^\prime\rangle- \langle\mathbf u^\prime.\nabla \omega_z^\prime\rangle, 
\label{eq:ozm}
\end{equation}
\begin{equation}
\um \cdot\nabla \uzm=\partial_z \frac\pmean\rho- \langle\mathbf u^\prime.\nabla u_z^\prime\rangle, 
\label{eq:uzm}
\end{equation}
\begin{equation}
\Delta \frac\pmean\rho=2\Omega\ozm-\mathcal \nabla\cdot\langle \mathbf u\cdot\nabla\mathbf u\rangle. \label{eq:pmean} 
\end{equation}
In (\ref{eq:pmean}),  $|\nabla\cdot\langle \mathbf u\cdot\nabla\mathbf u\rangle|/|\Omega\ozm|=\mathcal O(Ro)$, so for fast rotating turbulence ($Ro\ll1$), the average pressure is mostly governed by a geostrophic balance 
\begin{equation}
\frac\pmean\rho = 2\Omega\Delta^{-1}\ozm +\mathcal O(U^2Ro).
\label{eq:geo}
\end{equation}
Scaling arguments do not permit us to further simplify (\ref{eq:ozm}), (\ref{eq:uzm}). The reason is that since columnar structures are far longer than wide ($ l_z^I \gg l_\perp$), $z-$derivatives can be approximated as $\partial_z\sim (l_z^I)^{-1}$, implying that all terms in (\ref{eq:uzm}) are $\mathcal O (U^2/l_\perp)$ and all terms in (\ref{eq:ozm}) are $\mathcal O (U^2/l_\perp^2)$. The potential influence of fluctuations on the anisotropy of the average flow can, however, be analysed by experimentally evaluating the magnitude of all the terms in (\ref{eq:ozm}) and (\ref{eq:uzm}). Of particular interest are the last two terms in (\ref{eq:ozm}) and the last term in (\ref{eq:uzm}) as fluctuations and thus inertial waves, can only affect the average flow through them.

\section{Experimental methods} 
\label{sec:experiment}
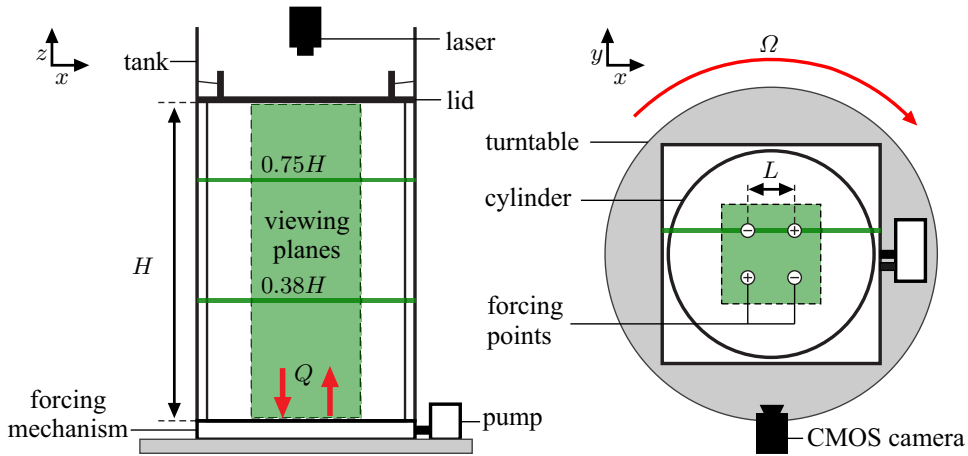
\begin{figure}
	\centerline{
		\includegraphics[scale=0.9]{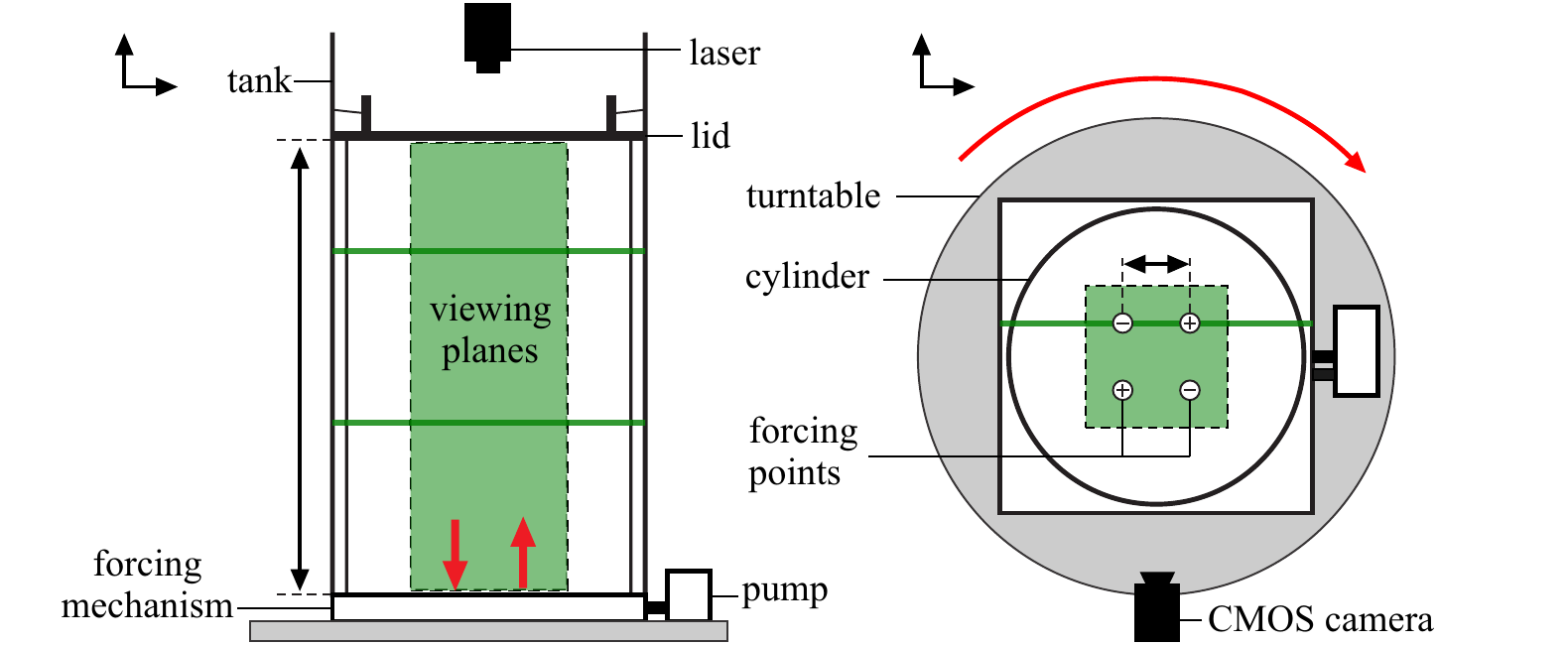}%
		\put(-372,141){\makebox(0,0)[r]{\strut{} $x$}}%
		\put(-380,150){\makebox(0,0)[r]{\strut{} $z$}}%
		\put(-162,141){\makebox(0,0)[r]{\strut{} $x$}}%
		\put(-171,150){\makebox(0,0)[r]{\strut{} $y$}}%
		\put(-340,70){\makebox(0,0)[r]{\strut{} $H$}}%
		\put(-105,106){\makebox(0,0)[r]{\strut{} $L$}}%
		\put(-105,153){\makebox(0,0)[r]{\strut{} $\Omega$}}%
		\put(-280,30){\makebox(0,0)[r]{\strut{} $Q$}}%
		\put(-275,63){\makebox(0,0)[r]{\strut{} $0.38H$}}%
		\put(-275,108){\makebox(0,0)[r]{\strut{} $0.75H$}}%
	}
	\caption{Side- and top-view sketch of experimental setup. Green regions and lines show areas and positions of PIV planes used during measurements. In top-view (+) refers to a source and (-) to a sink.}
	\label{fig1_setup}
\end{figure}
\subsection{Experimental apparatus}
The experimental setup consists of a rectangular tank (600 mm$\times$320 mm$\times$320 mm) fitted at the centre of a rotating turntable. The flow is forced by injecting and subtracting fluid through four holes (diameter  $d=1$ mm) located at the corners of a $L=53$ mm square in the bottom wall of the tank (figure \ref{fig1_setup}). All holes are connected to a peristaltic pump simultaneously injecting fluid through holes along one diagonal of the square and sucking fluid through the others, at the same constant flowrate $Q$ through each hole. This configuration was chosen to both cancel the injected mass flux, and to minimize the displacement of vortices forming along the subtraction and injection jets. A cylinder (height 400 mm, $\varnothing$ 300 mm) closed by an upper transparent lid placed inside the tank prevents free surface deformation and provides a viewing window for the optical measurements. The setup is spun up into solid body rotation at a rotation speed $\Omega$, before the pump is initiated. Prior to measurements, the flow is left to settle into a statistically steady state. Statistical steadiness was assessed through the convergence of statistical quantities, kept below 1\% in most cases.\\ 

The governing parameters are the Ekman number $E=\nu/2\Omega H^2$ and a forcing-based Reynolds number, $Re_Q=4Q/\pi\nu d$. They are independently controlled by $\Omega$ and $Q$. The flow intensity is measured \emph{a posteriori} by means an average-based and a fluctuations-based Rossby numbers $Ro=\langle |\mathbf u|\rangle_{\mathbf x t}/2\Omega L$ and $Ro^\prime=\langle |\mathbf u^{\prime2}|\rangle_{\mathbf x t}^{1/2}/2\Omega L$, built on time and space averages $\langle \cdot\rangle_{\mathbf x t}$ over the horizontal visualisation plane at $z=0.75H$. Experiments are performed over a range of parameters spanning $E=\{4.25, 5.67, 8.59, 17, 34\}\times10^{-5}$ and $3\times10^2\leq Re_Q\leq 1.5\times10^4$. In this range, the jets penetrating the flow are always turbulent. Velocity fields are measured with a 2D-PIV system: a laser sheet illuminates horizontal planes (HP) at $z=0.38H$ or $z=0.75H$, or a vertical plane (VP) aligned on a injection/subtraction pair. For visualisations in the HP, a 1.3MP CMOS camera records a 150 mm $\times$ 150 mm area centred on the tank at 30 fps. For the VP experiments, two cameras record an area of 400 mm $\times$150 mm at 60 fps along the tank. The smallest resolvable distance is 2.1 mm in all planes.\\

\subsection{Evaluating the average flow quantities}
\label{sec:upperbounds}
Evaluating $\langle\boldsymbol{\omega}^\prime\cdot\nabla u_z^\prime\rangle$, $\langle\mathbf{u'}\cdot\nabla\omega^\prime_z\rangle$ and $\langle\mathbf{u'}\cdot\nabla u^\prime_z\rangle$, requires us to calculate expressions such as $\partial_z \omega_z$ that are not directly accessible from 2D-PIV data. It is however possible to calculate rigorous upper bounds for these three-dimensional quantities, using the two-dimensional quantities accessible through plane PIV measurements only. 
We start by noticing that the symmetry of the forcing and the geometry imply that statistical properties are statistically invariant by rotation of $\pi/2$ followed by a symmetry with respect to any vertical plane equidistant from two fluid injection/subtraction points located on any one side of the square. Hence statistical properties in the $x$ and $y$ directions are identical, so that
\begin{equation}
|\langle \mathbf u^\prime\cdot \nabla u_z^\prime\rangle|=|\langle u_z^\prime\partial_zu_z^\prime\rangle|+2|\langle u_x^\prime\partial_xu_z^\prime\rangle|=|\langle \mathbf u^\prime\cdot \nabla u_z^\prime\rangle|^e~.
\end{equation}
Here the superscript $^e$ refers to the estimates for the three-dimensional terms that we built out of 
quantities that we actually measured with the 2D PIV system. An upper bound estimate is obtained for $|\langle\boldsymbol{\omega}^\prime\cdot\nabla u_z^\prime\rangle|$, using Schwartz's inequality to bound the average of products with the product of averages, Minkowsky's inequality to handle sums, and again, statistical equivalence of directions $x$ and $y$:
\begin{equation}
|\langle\boldsymbol{\omega}^\prime\cdot\nabla u_z^\prime\rangle|\leq 
2\langle| \omega_y^\prime|^2\rangle^{1/2}\langle| \partial_x u_z^\prime|\rangle^{1/2}+\langle|\omega_z^\prime|^2\rangle^{1/2}\langle|\partial_zu_z^\prime|^2\rangle^{1/2}=|\langle\boldsymbol{\omega}^\prime\cdot\nabla u_z^\prime\rangle|^e.
\end{equation}
In the resulting estimate, all terms are evaluated from VP-PIV except $\langle|\omega_z^\prime|^2|\rangle^{1/2}$, which is obtained from HP-PIV. 
Using a similar approach, an upper bound estimate is obtained for the reminding three-dimensional term as
\begin{equation}
|\langle\mathbf u^\prime\cdot\nabla \omega_z^\prime\rangle|\leq |\langle u_x^\prime\partial_x \omega_z^\prime\rangle+\langle u_y^\prime\partial_y \omega_z^\prime\rangle| + \langle |u_z^\prime|^2\rangle^{1/2} \langle|\partial_z \omega_z^\prime|^2\rangle^{1/2}=|\langle\mathbf u^\prime\cdot\nabla \omega_z^\prime\rangle|^e.
\end{equation}
Additionally, contributions from inertial waves to these terms are estimated by filtering out velocity and vorticity components whose frequency exceeds the maximum possible frequency of inertial waves, $2\Omega$ \citep{greenspan}. An upper bound for the contribution of inertial waves is obtained by treating all remaining fluctuations in terms filtered in this way as if they were inertial waves. As such, $\langle|\partial_z \omega_z^\prime|^2\rangle^{1/2}$ is estimated by replacing $\partial_z$ by an upper estimate $V_I(H)\partial_t$, where $V_I(H)$ is the fastest inertial wave group velocity, \emph{i.e.} that associated to the largest possible scale in the vessel, $H$.\\
To find an upper bound for the contribution of inertial waves to estimates for the nonlinear terms derived in section \ref{sec:upperbounds}, each of the measured velocity and fields are split in their two components $f>2\Omega$ and $f<2\Omega$. As such, nonlinear terms express as four components, only one of which is exclusively built from wave-free velocity contributions. The remaining three components form the upper bound for the contribution of inertial waves to the nonlinear term. Spatial derivatives are calculated by means of a 4th order centred finite difference scheme.\\
We stress again here that the derivation of these upper bounds requires no simplifying assumption, other than assuming sufficient statistical convergence of our dataset to achieve statistical equivalence of properties along $x$ and $y$. As such these upper bounds require no validation, as long as they are handled as such. In particular, to show that the actual three-dimensional terms are negligible compared to the contribution of the Coriolis force, it is sufficient to show that the 2D upper bounds are, without further validation. The flip side of this approach is that it is difficult to estimate how close estimates are from the actual values. Consequently, even if the estimates are not found significantly smaller than the contribution of the Coriolis force, the actual three-dimensional quantities may still be.\\

\section{Results}
\label{sec:results}
For the range of parameters we consider, turbulent jets form above the two injection/subtraction points and feed a small turbulent patch dominated by inertia rather than by the Coriolis force. This patch extends to a critical height $h_p$ such that the local Rossby number at $h_p$ reaches unity. A similar patch exists in rotating grid turbulence \citep{dickinson83} and in turbulent convective plumes under the effect of rotation \citep{maxworthy94}. A more detailed analysis of this region can be found in \citet{bpt2020_jfm_T}.
Columnar structures develop above the patch where $z>h_p$. The lower limit ($Ro<10^{-2}$) coincides with a regime of weaker turbulence, dominated by a quadrupole of robust columnar structures, reminiscent of the 2D flow solutions identified by \citet{gallet2015_jfm} when rotating flows become two-dimensional. The quadrupole sits above the turbulent patch and is aligned with the 4 injection/subtraction holes. At higher $Ro$, the columns unlock from the forcing points and become subject to mutual advection, pairing and merging \citep{tabeling2002}.\\
\subsection{Columnar length scale in forced rotating turbulence}
\begin{figure}
	\centerline{
		\includegraphics[scale = 0.9]{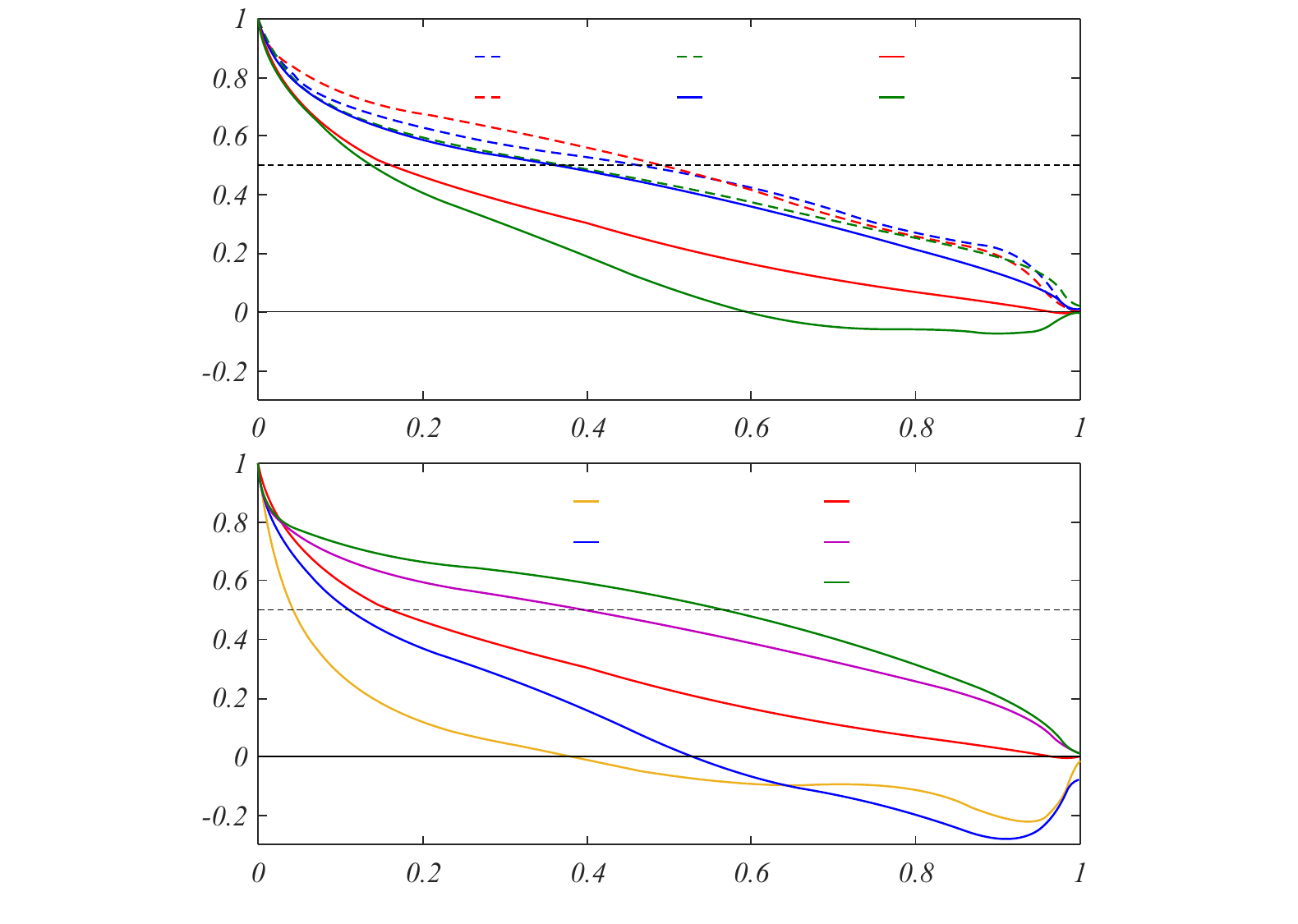}
		\put(-190,3){{\makebox(0,0)[r]{\strut{} $\delta_z / H$}}}%
		\put(-350,285){{\makebox(0,0)[r]{\strut{} $a)$}}}%
		\put(-265,268){{\makebox(0,0)[r]{\strut{} $Re_Q$}}}%
		\put(-225,274){{\makebox(0,0)[r]{\strut{} $2500$}}}%
		\put(-225,261){{\makebox(0,0)[r]{\strut{} $4000$}}}%
		\put(-162,274){{\makebox(0,0)[r]{\strut{} $6000$}}}%
		\put(-162,261){{\makebox(0,0)[r]{\strut{} $7500$}}}%
		\put(-100,274){{\makebox(0,0)[r]{\strut{} $9000$}}}%
		\put(-98,261){{\makebox(0,0)[r]{\strut{} $12000$}}}%
		\put(-355,240){\rotatebox{-270}{\makebox(0,0)[r]{\strut{} $\frac{C_{u_x}(\delta z)}{C_0}$}}}%
		\put(-350,145){{\makebox(0,0)[r]{\strut{} $b)$}}}%
		\put(-238,127){{\makebox(0,0)[r]{\strut{} $E$}}}%
		\put(-168,133){{\makebox(0,0)[r]{\strut{} $34.0\times10^{-5}$}}}%
		\put(-168,120){{\makebox(0,0)[r]{\strut{} $17.0\times10^{-5}$}}}%
		\put(-90,133){{\makebox(0,0)[r]{\strut{} $8.50\times10^{-5}$}}}%
		\put(-90,120){{\makebox(0,0)[r]{\strut{} $5.67\times10^{-5}$}}}%
		\put(-90,107){{\makebox(0,0)[r]{\strut{} $4.25\times10^{-5}$}}}%
		\put(-355,100){\rotatebox{-270}{\makebox(0,0)[r]{\strut{} $\frac{C_{u_x}(\delta z)}{C_0}$}}}%
	}
	\caption{Normalized two-point velocity correlations $C_{u_x}(\delta z)$ based on separation distance $\delta z$. a) $C_{u_x}$ across various $Re_Q$ for $E=8.50\times10^{-5}$. b) $C_{u_x}$ for various $E$ at $Re_Q=9000$. Dashed black line represent threshold value $\beta$ used to calculate correlation length $l_z$ using (\ref{eq2_LZ}) across all experiments.}
	\label{fig2}
\end{figure}
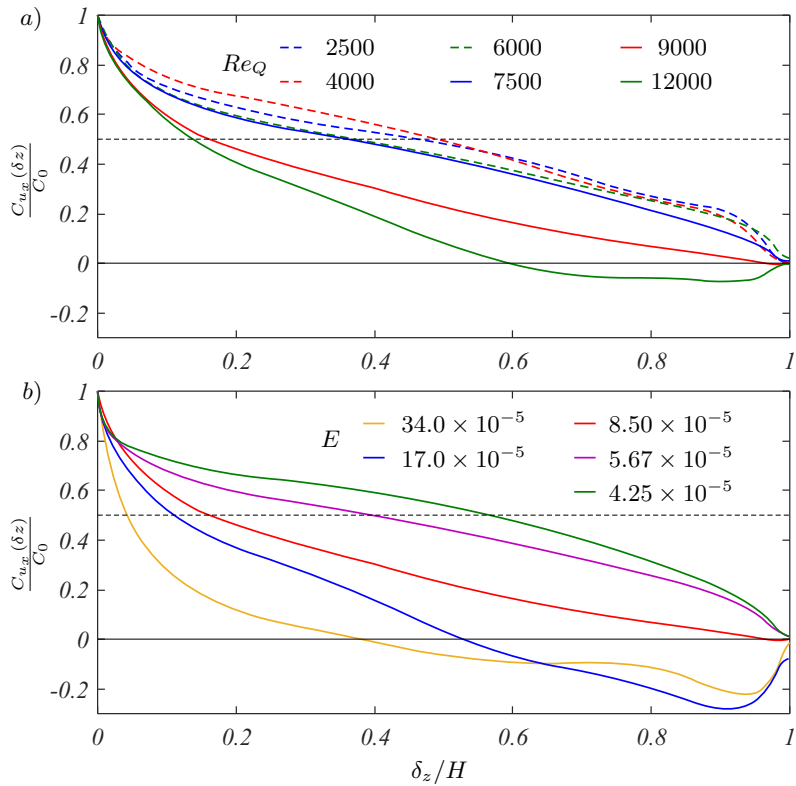
We start by confronting the scalings for the vertical length scales (\ref{eq:lz_inertia}) to experimental data. We build two vertical length scales $l_z$ and $l_z^\prime$ from experiments along the vertical plane using two-point velocity correlations $C_{u_x}(\delta z)$ and $C_{u_x^\prime}(\delta z)$, respectively calculated from the full velocity field $u_x$ or its fluctuating part $u_x^\prime$ \citep{aujogue18_jfm}, using:
\begin{equation}
C_{u_x}(\delta z) = \langle{\int_{A}u_x(x,z+\delta z)u_x(x,z)dxdz}\rangle_{t}~.
\label{eq1_LZ}
\end{equation}
Here $A$ is the area of the flow field captured by the experiments along the vertical plane, $\langle \cdot\rangle_{t}$ is a temporal average and $\delta z$ is the separation between two points along the $z-$axis. These correlations are normalized by a constant $C_0= C_{u_x}(0).$ Figure \ref{fig2} shows $C_{u_x}(\delta z)$ across various control parameters. Similar behaviour is seen for $C_{u_x^\prime}(\delta z)$. For $\delta z/H>0.9$ $C_{u_x}(\delta z)$ tends to zero. This is caused by the upper boundary of the tank. 

The characteristic length scales are generally determined by finding the first zero in $C_{u_x}(\delta z)$. Figure \ref{fig2} shows that in practice $C_{u_x}(\delta z)$ does not necessarily fully decorrelate over $h_p\leq z\leq H$. 
Hence, following \citet{staplehurst08}, $l_z$ and $l_z^\prime$ are derived using an arbitrary threshold value $\beta$. At separation distance $Z$, when $C_{u_x}(z)/C_0=\beta$, $l_z$ is defined using: 
\begin{equation}
\ell_z = \int_{0}^{Z}{{C_{u_x}(z)}\over{C_0}} dz~.
\label{eq2_LZ}
\end{equation}
As can be seen in figure \ref{fig2}a and b the behaviour of correlations shift drastically across both $E$ and $Re_Q$. Therefore we chose $\beta=0.5$ so as to be able to apply the same method across all experiments.\\
\begin{figure}
	\centerline{
		\includegraphics[scale = 0.9]{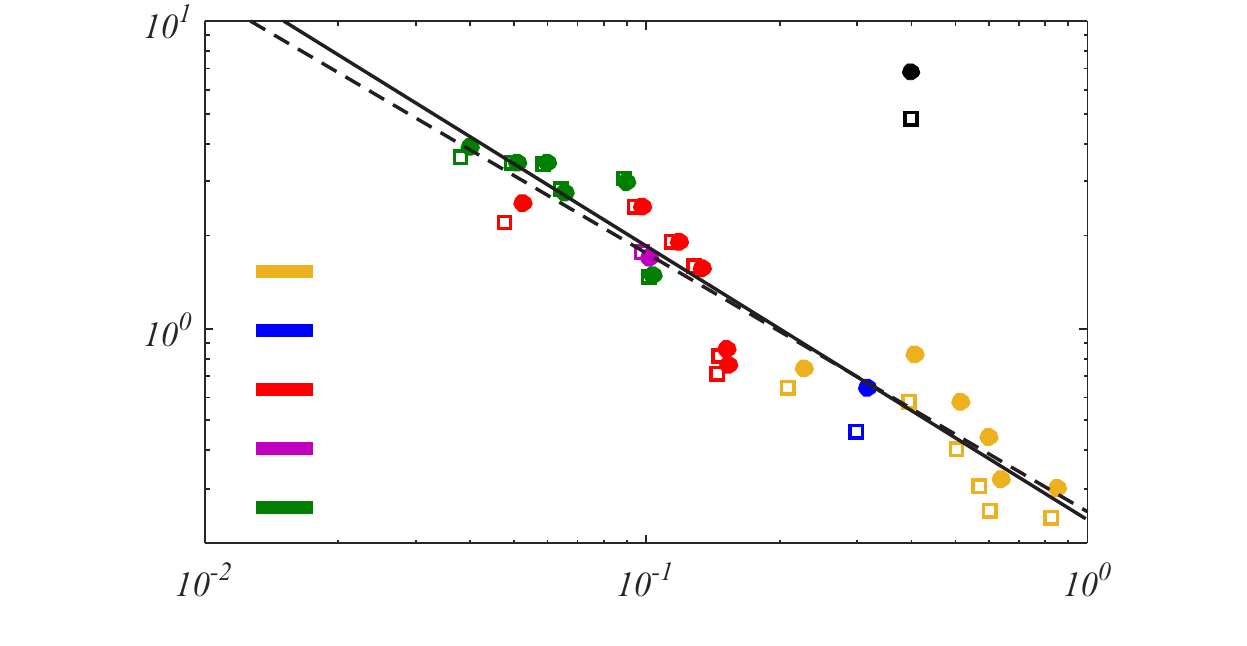}
		\put(-70,150){{\makebox(0,0)[r]{\strut{} $l_z$}}}%
		\put(-70,136){{\makebox(0,0)[r]{\strut{} $l^\prime_z$}}}%
		\put(-160,5){{\makebox(0,0)[r]{\strut{} $Ro$}}}%
		\put(-215,108){{\makebox(0,0)[r]{\strut{} $E$}}}%
		\put(-190,96){{\makebox(0,0)[r]{\strut{} $34.0\times10^{-5}$}}}%
		\put(-190,80){{\makebox(0,0)[r]{\strut{} $17.0\times10^{-5}$}}}%
		\put(-190,65){{\makebox(0,0)[r]{\strut{} $8.50\times10^{-5}$}}}%
		\put(-190,50){{\makebox(0,0)[r]{\strut{} $5.67\times10^{-5}$}}}%
		\put(-190,35){{\makebox(0,0)[r]{\strut{} $4.25\times10^{-5}$}}}%
		\put(-170,150){{\makebox(0,0)[r]{\strut{} $0.24~Ro^{-0.9}$}}}%
		\put(-213,127){{\makebox(0,0)[r]{\strut{} $0.26~Ro^{-0.8}$}}}%
		\put(-300,100){\rotatebox{-270}{\makebox(0,0)[r]{\strut{} $l/L$}}}%
	}
	\caption{Columnar structure length $l_z$ based on $u_x$ and $l^\prime_z$ based on $u^\prime_z$ normalised by $L$. $Ro$ based on $|\mathbf u|$ and $|\mathbf u^\prime|$ respectively. Solid and dashed black line show a fit of $l_z$ and $l^\prime_z$ data respectively. }
	\label{fig3_LZ}
\end{figure}
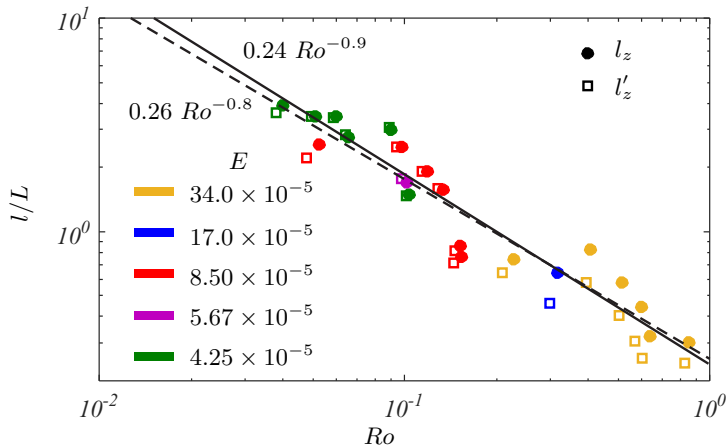
Due to the limited space across which the correlations can be applied ($h_p\leq z\leq H$) and the application of the arbitrary threshold there is scattering in the data for $\l_z$ and $\l_z'$, as seen in figure \ref{fig3_LZ}. Nevertheless, both $\l_z$ and $\l_z'$ closely follow the $l_z\sim L Ro^{-1}$ scaling from (\ref{eq:lz_inertia}). This confirms that columns above the turbulent patch form under the combined influence of the Coriolis forces and inertia. Inertia may however be associated to the average flow or to fluctuations, which in turn may be either random or driven by inertial waves. Evaluating the relative importance of the terms in equations (\ref{eq:ozm}$-$\ref{eq:pmean}) shall therefore highlight flow regimes where inertial waves are active.\\ 

\subsection{Separating inertial waves from turbulent fluctuations}
\label{sec:iwsep}
\begin{figure}
	\centerline{
		\includegraphics[scale = 0.9]{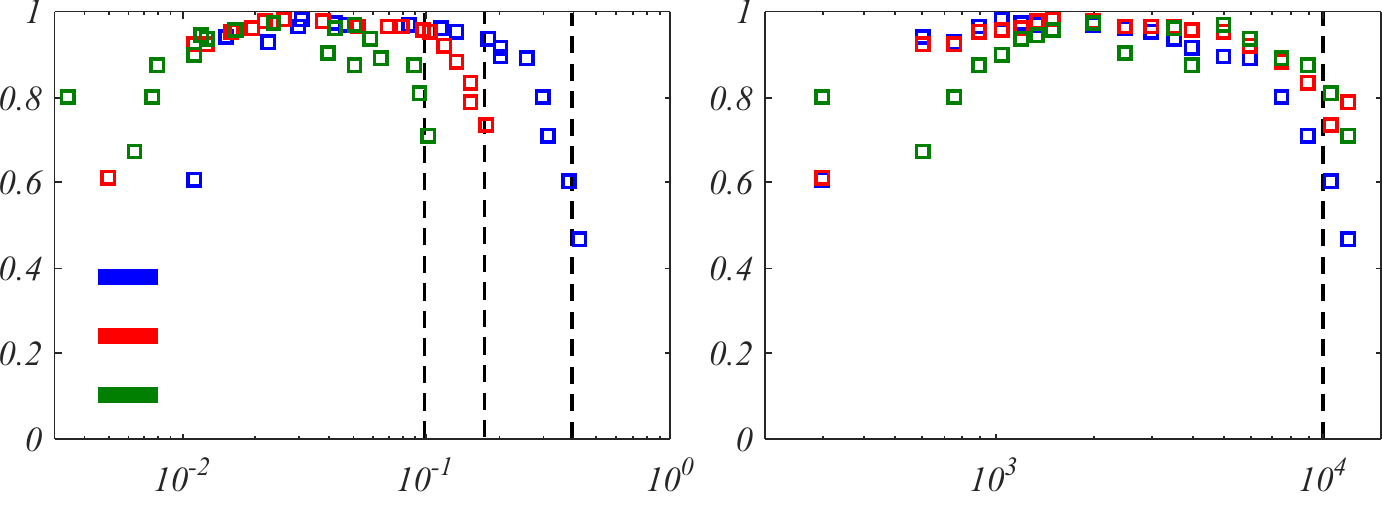}
		\put(-359,132){\makebox(0,0)[r]{\strut{} \footnotesize $a)$}}%
		\put(-172,132){\makebox(0,0)[r]{\strut{} \footnotesize $b)$}}%
		\put(-255,0){\makebox(0,0)[r]{\strut{} \footnotesize $Ro^\prime$}}%
		\put(-75,0){\makebox(0,0)[r]{\strut{} \footnotesize $Re_Q$}}%
		\put(-290,76){{\makebox(0,0)[r]{\strut{} $E$}}}%
		\put(-265,64){{\makebox(0,0)[r]{\strut{} $17.0\times10^{-5}$}}}%
		\put(-265,48){{\makebox(0,0)[r]{\strut{} $8.50\times10^{-5}$}}}%
		\put(-265,33){{\makebox(0,0)[r]{\strut{} $4.25\times10^{-5}$}}}%
		\put(-370,95){\rotatebox{-270}{\makebox(0,0)[r]{\strut{} $E^\prime_{IW}/E^\prime$}}}%
	}
	\caption{Upper bound of energy carried by inertial wave fluctuations to energy carried by the fluctuations \emph{vs.} $a)$ $Ro^\prime$ and $b)$ $Re_Q$ for various $E$ at $z=0.75H$.}
	\label{fig4}
\end{figure}

\begin{figure}
	\centerline{
		\includegraphics[scale = 0.9]{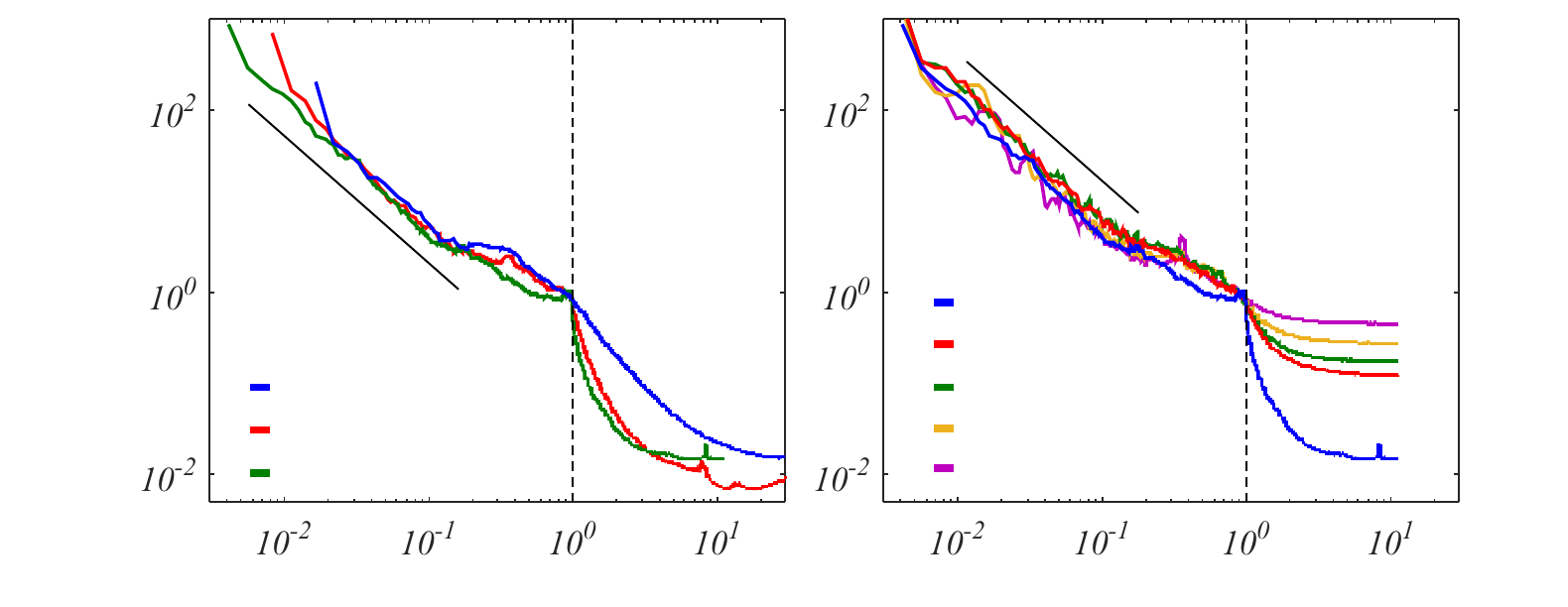}
		\put(-270,4){\makebox(0,0)[r]{\strut{} $f/2\Omega$}}%
		\put(-95,4){\makebox(0,0)[r]{\strut{} $f/2\Omega$}}%
		\put(-365,150){\makebox(0,0)[r]{\strut{} $a)$}}%
		\put(-190,150){\makebox(0,0)[r]{\strut{} $b)$}}%
		\put(-90,130){\makebox(0,0)[r]{\strut{} $\propto(\frac{f}{2\Omega})^{-1.39}$}}%
		\put(-130,90){\makebox(0,0)[r]{\strut{} $Re_Q$}}%
		\put(-130,78){\makebox(0,0)[r]{\strut{} $2000$}}%
		\put(-130,67){\makebox(0,0)[r]{\strut{} $4000$}}%
		\put(-130,56){\makebox(0,0)[r]{\strut{} $6000$}}%
		\put(-130,45){\makebox(0,0)[r]{\strut{} $9000$}}%
		\put(-128,34){\makebox(0,0)[r]{\strut{} $12000$}}%
		\put(-300,85){\makebox(0,0)[r]{\strut{} $\propto(\frac{f}{2\Omega})^{-1.39}$}}%
		\put(-305,68){\makebox(0,0)[r]{\strut{} $E$}}%
		\put(-285,56){\makebox(0,0)[r]{\strut{} $17.0\times10^{-5}$}}%
		\put(-285,45){\makebox(0,0)[r]{\strut{} $8.50\times10^{-5}$}}%
		\put(-285,34){\makebox(0,0)[r]{\strut{} $4.25\times10^{-5}$}}%
		\put(-385,120){\rotatebox{-270}{\makebox(0,0)[r]{\strut{}$\langle E(f)\rangle/\langle E(2\Omega)\rangle$}}}%
	}
	\caption{Power spectra normalized by $\langle E(2\Omega)\rangle$ at a) $Re_Q=2000$ with varying $E$ and b) $E=4.25\times10^{-5}$ with varying $Re_Q$.} 
	\label{fig5}
\end{figure}
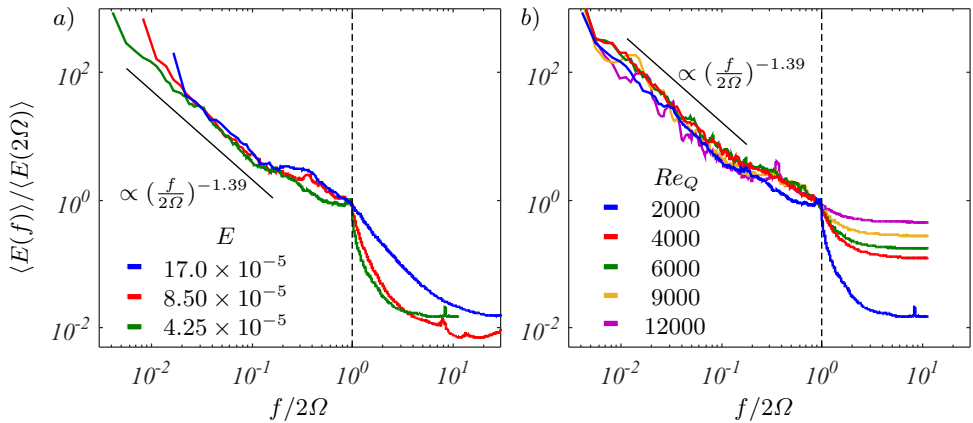

\begin{figure}
	\centerline{
		\includegraphics[scale = 0.9]{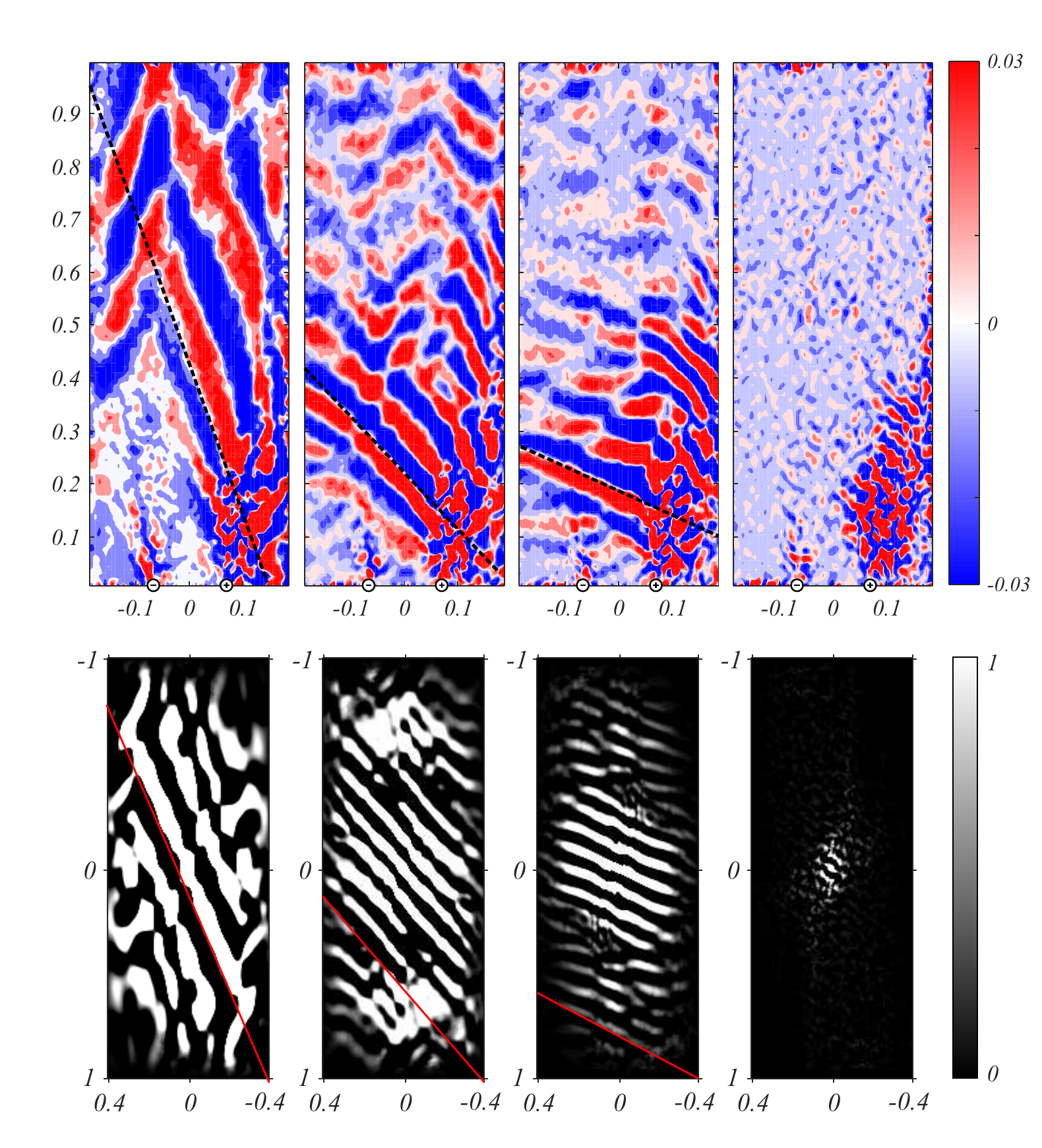}
		\put(-326,5){\makebox(0,0)[r]{\strut{} $dx/H$}}%
		\put(-241,5){\makebox(0,0)[r]{\strut{} $dx/H$}}%
		\put(-156,5){\makebox(0,0)[r]{\strut{} $dx/H$}}%
		\put(-71,5){\makebox(0,0)[r]{\strut{} $dx/H$}}%
		\put(-390,125){\rotatebox{-270}{\makebox(0,0)[r]{\strut{}$dz/H$}}}%
		\put(-15,100){\rotatebox{-90}{\makebox(0,0)[r]{\strut{}$C_{xz}$}}}%
		\put(-375,201){\makebox(0,0)[r]{\strut{}\footnotesize $e)$}}%
		\put(-290,201){\makebox(0,0)[r]{\strut{}\footnotesize $f)$}}%
		\put(-205,201){\makebox(0,0)[r]{\strut{}\footnotesize $g)$}}%
		\put(-120,201){\makebox(0,0)[r]{\strut{}\footnotesize $h)$}}%
		\put(-328,201){\makebox(0,0)[r]{\strut{} $x/H$}}%
		\put(-243,201){\makebox(0,0)[r]{\strut{} $x/H$}}%
		\put(-158,201){\makebox(0,0)[r]{\strut{} $x/H$}}%
		\put(-73,201){\makebox(0,0)[r]{\strut{} $x/H$}}%
		\put(-310,433){\makebox(0,0)[r]{\strut{}\footnotesize $a)~~f/2\Omega=0.34$}}%
		\put(-225,433){\makebox(0,0)[r]{\strut{}\footnotesize $b)~~f/2\Omega=0.64$}}%
		\put(-140,433){\makebox(0,0)[r]{\strut{}\footnotesize $c)~~f/2\Omega=0.87$}}%
		\put(-55,433){\makebox(0,0)[r]{\strut{}\footnotesize $d)~~f/2\Omega=1.5$}}%
		\put(-400,336){\rotatebox{-270}{\makebox(0,0)[r]{\strut{}$z/H$}}}%
		\put(-5,306){\rotatebox{-90}{\makebox(0,0)[r]{\strut{}$\omega_y^\prime(s^{-1})$}}}%
	}
	\caption{Top: Filtered vorticity field $\omega_y^\prime$ at various frequencies $f$. Dashed black line represent the angle of propagation $\theta$ predicted by the dispersion relation for inertial waves. (+) and (-) represent the approximate location of the source and sink, respectively. When $f<2\Omega$ wave-like patterns are found. Bottom: two-dimensional cross-correlations of $u_x^\prime$ filtered at frequencies $f$ for $Re_Q=1200$ and $E=4.25\times10^{-5}$. Red lines represent the propagation angle $\theta$ predicted by dispersion relation for inertial waves.} 
	\label{fig6}
\end{figure}

\begin{figure}
	\centerline{
		\includegraphics[scale = 0.9]{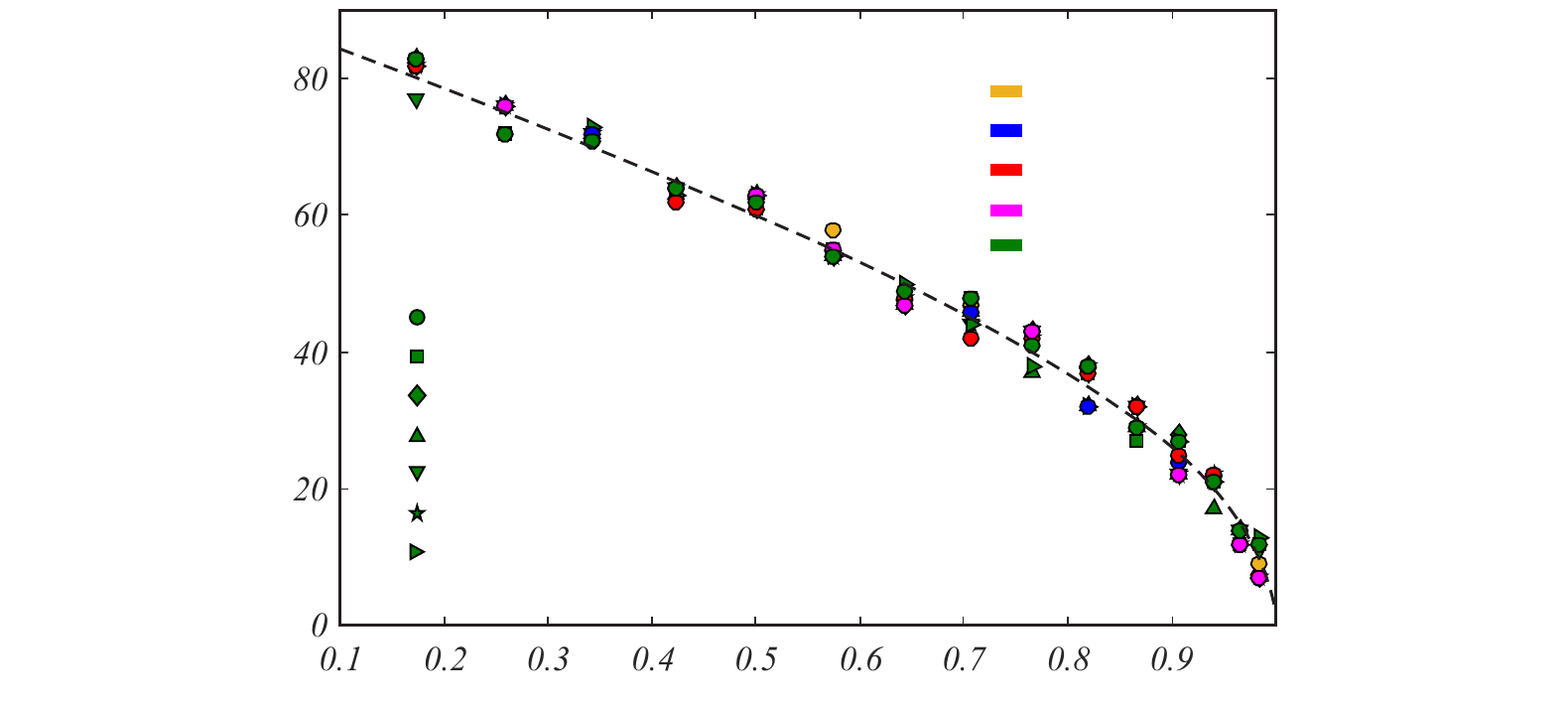}
		\put(-190,0){\makebox(0,0)[r]{\strut{} $f/2\Omega$}}%
		\put(-110,172){\makebox(0,0)[r]{\strut{} $E$}}%
		\put(-90,160){\makebox(0,0)[r]{\strut{} $34.0\times10^{-5}$}}%
		\put(-90,150){\makebox(0,0)[r]{\strut{} $17.0\times10^{-5}$}}%
		\put(-90,140){\makebox(0,0)[r]{\strut{} $8.50\times10^{-5}$}}%
		\put(-90,130){\makebox(0,0)[r]{\strut{} $5.67\times10^{-5}$}}%
		\put(-90,120){\makebox(0,0)[r]{\strut{} $4.25\times10^{-5}$}}%
		\put(-272,112){\makebox(0,0)[r]{\strut{} $Re_Q$}}%
		\put(-274,100){\makebox(0,0)[r]{\strut{} $600$}}%
		\put(-272,90){\makebox(0,0)[r]{\strut{} $1200$}}%
		\put(-272,80){\makebox(0,0)[r]{\strut{} $2500$}}%
		\put(-272,70){\makebox(0,0)[r]{\strut{} $4000$}}%
		\put(-272,60){\makebox(0,0)[r]{\strut{} $6000$}}%
		\put(-272,50){\makebox(0,0)[r]{\strut{} $7500$}}%
		\put(-270,40){\makebox(0,0)[r]{\strut{} $12000$}}%
		\put(-350,130){\rotatebox{-270}{\makebox(0,0)[r]{\strut{}$\theta~(degrees)$}}}%
	}
	\caption{Propagation angle $\theta$ of wave-like patterns identified from vorticity fields $\omega_y^\prime$ filtered at frequency $f$. Dashed black line represent the dispersion relation for inertial waves \citet{greenspan}} 
	\label{fig7}
\end{figure}

\begin{figure}
	\centerline{
		\includegraphics[scale = 0.85]{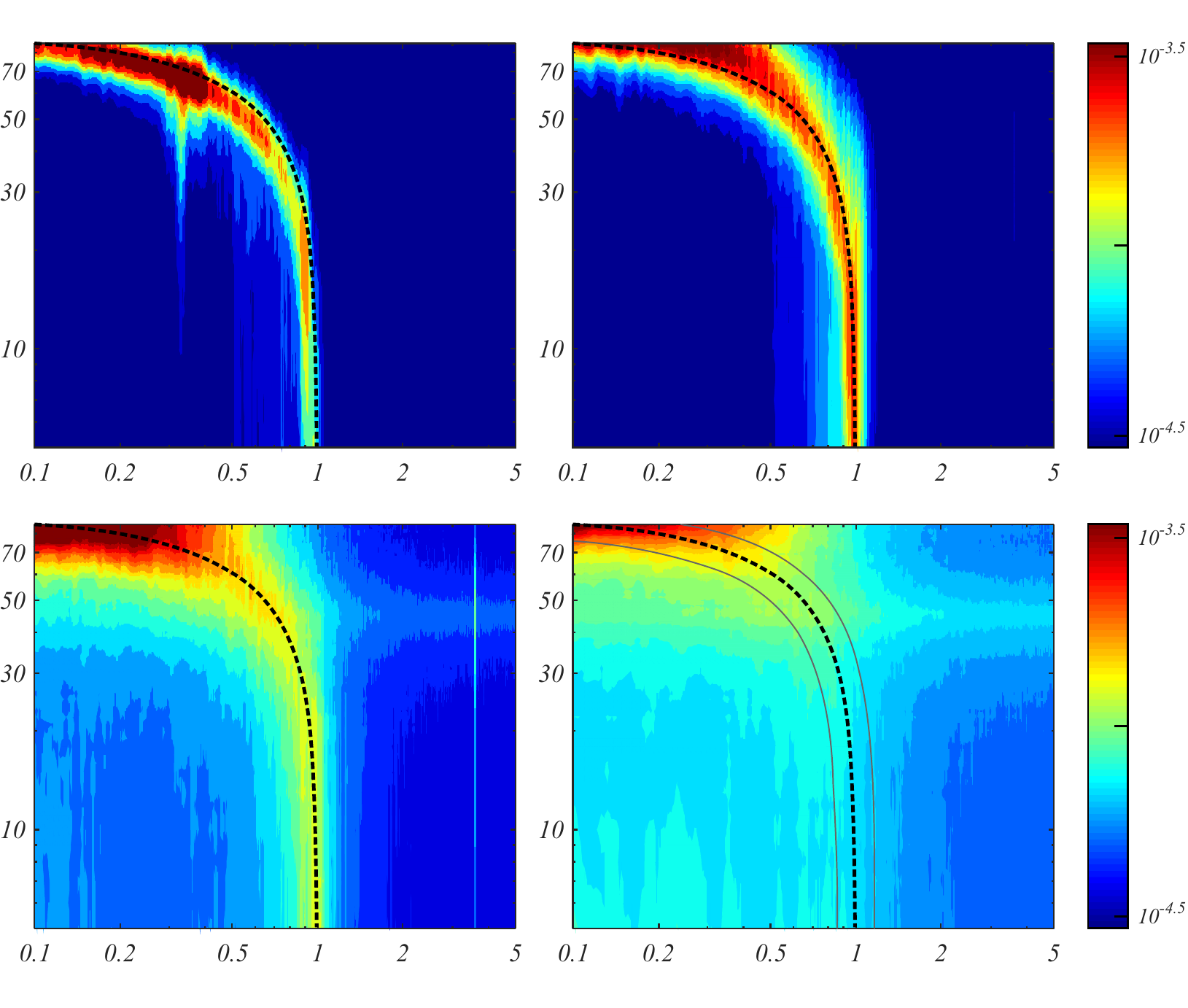}
		\put(-260,330){\makebox(0,0)[r]{\strut{} \footnotesize $a)~Re_Q~600,~Ro~0.006$}}%
		\put(-80,330){\makebox(0,0)[r]{\strut{} \footnotesize $b)~Re_Q~1200,~Ro~0.014$}}%
		\put(-260,168){\makebox(0,0)[r]{\strut{} \footnotesize $c)~Re_Q~2500,~Ro~0.036$}}%
		\put(-80,168){\makebox(0,0)[r]{\strut{} \footnotesize $d)~Re_Q~7500,~Ro~0.096$}}%
		\put(-320,5){\makebox(0,0)[r]{\strut{} \footnotesize $f/2\Omega$}}%
		\put(-120,5){\makebox(0,0)[r]{\strut{} \footnotesize $f/2\Omega$}}%
		\put(-405,105){\rotatebox{-270}{\makebox(0,0)[r]{\strut{} $\theta~(deg.)$}}}%
		\put(-405,267){\rotatebox{-270}{\makebox(0,0)[r]{\strut{} $\theta~(deg.)$}}}%
		\put(-5,235){\rotatebox{-90}{\makebox(0,0)[r]{\strut{} $E^\prime(\theta,f)/U_0^2$}}}%
		\put(-5,73){\rotatebox{-90}{\makebox(0,0)[r]{\strut{} $E^\prime(\theta,f)/U_0^2$}}}%
	}
	\caption{Energy density spectra $E^\prime(\theta,f)$ at $E=4.25\times10^{5}$ for various $Re_Q$. Dashed black line shows the dispersion relation for inertial waves. Grey lines in (d) show the narrow band considered to calculate $E_{band}^\prime$.}
	\label{fig8}
\end{figure}

\begin{figure}
	\centerline{
		\includegraphics[scale = 0.85]{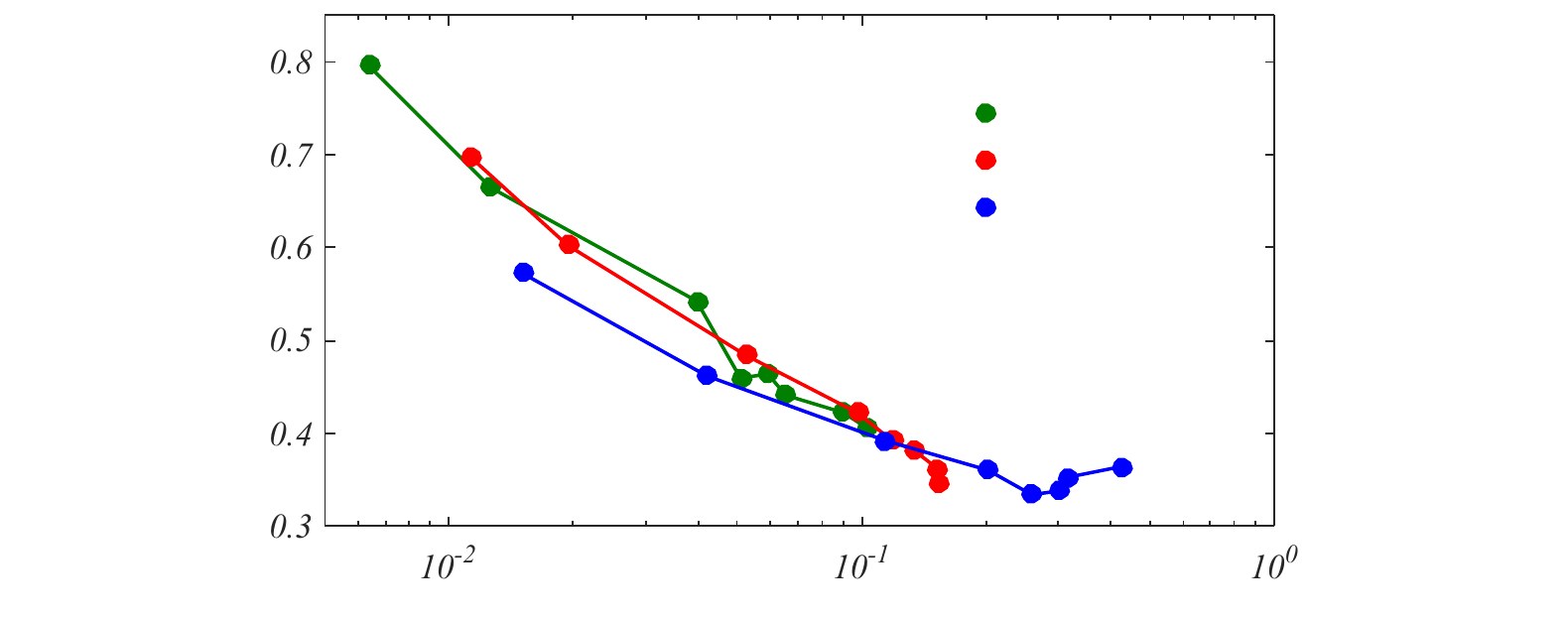}
		\put(-180,2){\makebox(0,0)[r]{\strut{} \footnotesize $Ro$}}%
		\put(-115,136){\makebox(0,0)[r]{\strut{} \footnotesize $E$}}%
		\put(-90,124){\makebox(0,0)[r]{\strut{} \footnotesize $4.25\times10^{-5}$}}%
		\put(-90,112){\makebox(0,0)[r]{\strut{} \footnotesize $8.50\times10^{-5}$}}%
		\put(-90,100){\makebox(0,0)[r]{\strut{} \footnotesize $17.0\times10^{-5}$}}%
		\put(-335,105){\rotatebox{-270}{\makebox(0,0)[r]{\strut{} $E_{band}^\prime/E^\prime_{\rm IW}$}}}%
	}
	\caption{Ratio $E_{band}^\prime/E^\prime_{\rm IW}$ as function of $Ro$ showing the efficiency of the filtering method applied.}
	\label{fig9}
\end{figure}
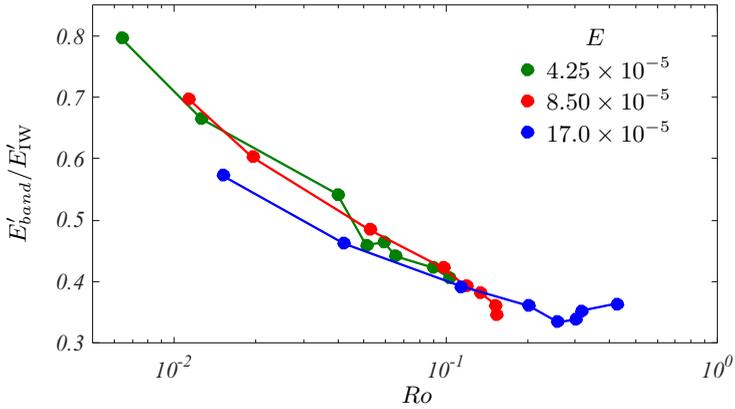

To this end, we first need to distinguish random turbulent fluctuations from inertial waves. This is done by splitting the turbulent energy spectrum into fluctuations of frequency $f$ greater than the maximum frequency of inertial waves $2\Omega$ (subscript FT) \citet{greenspan}, and fluctuations of frequency $f<2\Omega$, which may result from inertial waves or from random turbulence (subscript IW). The ratio of the total energy contained in the lower part of the spectrum $E^\prime_{\rm IW}$ to the total energy $E^\prime$ provides an upper bound for the fraction of the turbulent kinetic energy carried by inertial waves. Though global, this approach is similar to \citet{campagne15_pre}'s scale-dependent disentanglement method. 
The frequency spectra in figure \ref{fig4} show that most of the fluctuations' kinetic energy lies within the spectral range of inertial waves provided $Ro\gtrsim10^{-2}$ and $Re_Q<4000$. The sharp drop of energy in the spectra precisely at $f=2\Omega$ (figure \ref{fig5}a) suggests that the ratio $E^\prime_{\rm IW}/E^\prime$ reflects the relative importance of inertial waves, at least to some extent.
The absence of inertial waves in the higher range of either $Ro$ or $Re_Q$, reflects their disruption by random turbulence. In freely decaying turbulence, this phenomenon is controlled by the ratio between inertia and the Coriolis force, and takes place at $Ro^\prime\gtrsim0.4$ \citep{staplehurst08}. Here, inertial waves vanish for $Re_Q\gtrsim10^4$, independently of the intensity of the Coriolis force, most likely on the grounds that both the inertial waves and the inertia that disrupt them are driven by fluctuations in the turbulence patch whose intensity is entirely controlled by inertia.\\
\cite{dileoni14_pof,yarom14_nat} made similar observation of a sprectral energy drop 
at $f=2\Omega$, down to lower values of $E$. These authors also found that fluctuations in the range $f<2\Omega$ obeyed a scaling law $E(f)\propto f^{-1.35}$ consistent with the $E(f)\sim(f/2\Omega)^{-1.39}$ which we found in the lower frequency range of that region. Unlike these earlier experiments, however, our spectra exhibit an intermediate region $0.2 \lesssim f/2\Omega\leq1$, where $E(f)$ does not clearly follow a power law. Since the Rossby number associated to fluctuations in this range is larger than at the lower frequencies, and that the effects of rotation are weaker here than in the regimes of lower Ekman number considered by \citet{yarom14_nat}, it is likely that this intermediate range of frequencies is significantly affected by inertial effects unlike the range $f/2 \lesssim 0.2$, where the Coriolis force dominates.\\ 
Next, to better assess how well the $2\Omega$ frequency separates inertial waves form random fluctuations, we first analyse the flow patterns corresponding to frequencies respectively lower and higher than $2\Omega$, for $10^{-2}\leq Ro \leq10^{-1}$. Frequency-specific flow patterns are obtained by applying a pass-band filter centred on a given frequency $f$, to the time-dependent field $\mathbf (u_x^\prime,u_z^\prime)$, followed by phase-averaging, as in \citet{cortet10_pof}. The bandwidth of the filter $df$ was kept at constant range of $df/2\Omega=10^{-2}$.
For $f<2\Omega$, the resulting field reveals individual inertial wave-packets being radiated from the turbulent patch (illustrated on figure \ref{fig6}-a-c). These waves propagate throughout the flow field at an angle $\theta$, reflecting of the walls. Since these patterns are detected far enough from the wall, they are nearly axisymmetric on average, so the chevron patterns observed on figure \ref{fig6}-a-c are in fact the two-dimensional signature of the three-dimensional cones observed by \citep{duran13_pre}.
When $f>2\Omega$, by contrast, wave-like patterns give way to random fluctuations that remain mostly localised near the turbulent patch (figure \ref{fig6}-d). The propagation angle $\theta$ of the patterns is found by seeking the maximum of two-dimensional spatial cross correlations of the frequency filtered velocity components $(u_x^\prime,u_z^\prime)$, averaged over time. Figure \ref{fig6}-e-g show the correlation patterns for $u_x^\prime$, similar patterns are seen for $u_z^\prime$. This technique reveals patterns with a well defined propagation angle when $f<2\Omega$. When $f>2\Omega$, by contrast, signals only remain well-correlated around the origin, confirming that no inertial waves are present in this regime. The relation $\theta(f)$ obtained in this manner is represented on figure  \ref{fig7}, and found in very good agreement with the dispersion relation of inertial waves (\ref{eq:iw_disp}). This confirms that inertial waves are confined in the $f<2\Omega$ range and that they carry a significant fraction of the fluctuations' energy in this range. Interestingly, the presence of mean advection in the z−direction does no incur any Doppler-shift.\\
\\
Finally, to quantify relevance of the upper bounds  for the nonlinear terms obtained by filtering out the contribution of fluctuations of frequency larger than $2\Omega$, we now need to quantify the fraction of energy carried by inertial waves in the range $f<2\Omega$. We start with the spectral energy distribution of the fluctuations, represented in a $(k,\theta)$ plane in figure \ref{fig8}. These spectra are derived using techniques similar to \citet{yarom14_nat}.
In this representation, inertial waves are located near a line defined by their dispersion relation (\ref{eq:iw_disp}). Indeed, most of the energy is located there in the limit of low Ekman and Rossby numbers while the lower part of the graph is progressively contaminated as energy of frequency lower than $2\Omega$ is increasingly carried by random fluctuations. Interestingly, only a very small amount of energy is transferred to evanescent waves of frequency greater than $2\Omega$ at low $Ro$. The ratio of energy lying within 10\% of the dispersion wave (\emph{i.e.} $|f-f_{IW}(\theta)|/f<10\%$), where $f_{IW}(\theta)$ is the frequency of inertial wave propagating at angle $\theta$, is represented on figure \ref{fig9}. As expected, the upper bound captures a greater share of inertial waves in the $Ro\rightarrow0$ limit, and indeed a fair fraction of them up (up to 80\%, down to 30\%) for most of the values of $Ro$ investigated. This percentage is, however, only indicative, as it depends directly on how close to the exact dispersion relation fluctuations have to lie to be counted as inertial waves. Nevertheless, the fact remains that the simple filtering technique we employ not only provides a rigorous upper bound for the contribution of inertial waves but also a relevant one in the $Ro\rightarrow0$ limit. As such whenever quantities involving these bounds vanish in this limit, they provide unequivocal evidence that the contribution of inertial waves do. The flip side of using rigorous upper bounds is that they are not indicative of the behaviour of inertial waves for $Ro\gtrsim0.1$, and this shall have to be kept in mind when interpreting our results.\\
\\
%
\subsection{Contributions of inertial waves to anisotropy}
We are now in a position to estimate the nature and the magnitude of the contribution to anisotropy within the vertical plane due to fluctuations arising from inertial waves in (\ref{eq:ozm}-\ref{eq:pmean}). From figure \ref{fig10_fluc}a, contributions to the inertial terms in (\ref{eq:ozm}-\ref{eq:pmean}) are almost strictly due to fluctuations arising from inertial waves at $z=0.38H$. A similar trend is seen at $z=0.75H$. For $Ro\lesssim0.1$ we assume $|\langle \mathbf{u}^\prime\cdot\nabla u^\prime_z\rangle|^e\approx|\langle \mathbf{u}^\prime\cdot\nabla u^\prime_z \rangle|^e_{\rm IW}~,
|\langle \mathbf{u}^\prime\cdot\nabla\omega^\prime_z \rangle|^e\approx|\langle \mathbf{u}^\prime\cdot\nabla\omega^\prime_z \rangle|^e_{\rm IW}$ and
$|\langle \boldsymbol{\omega}^\prime\cdot\nabla u^\prime_z \rangle|^e\approx|\langle \boldsymbol{\omega}^\prime\cdot\nabla u^\prime_z \rangle|^e_{\rm IW}$.

From, figure \ref{fig10_fluc}b, the ratio $|\langle\mathbf u^\prime\cdot \nabla u_z^\prime\rangle|^e_{\rm IW}/|\um\cdot\nabla\uzm|$ scales as a positive power of $Ro$ (around $2$), both in the lower ($z=0.38H$) and upper ($z=0.75H$) parts of the flow for $Ro\lesssim0.15$. For $Ro\gtrsim2\times10^{-2}$, the estimate for fluctuations due to inertial waves is greater than inertia due to the average flow. In this regime, however, the estimate includes a significant contribution from random fluctuations (see section \ref{sec:iwsep}) and is therefore not indicative of the contribution of inertial waves.  For $Ro\lesssim 2\times10^{-2}$, by contrast, this ratio becomes lower than unity and in the limit $Ro\rightarrow 0$, fluctuations due to inertial waves cannot balance the pressure gradient in (\ref{eq:uzm}) so waves play no part in determining $\uzm$. Note that all scalings reported on figure \ref{fig10_fluc} are indicative only, and that while the quantities plotted in all four graphs follow the same exponent in $Ro$, there is no theoretical basis to expect that data obtained at different values of $E$ and different heights $H$ to collapse into the same curve. Consequently, it follows from the asymptotic behaviour in the limit $Ro\rightarrow0$, and from (\ref{eq:uzm}) and (\ref{eq:geo}) that
\begin{eqnarray}
\um \cdot \nabla \uzm&=&  2\Omega\Delta^{-1}\partial_z\ozm +{\rm FT}+\mathcal O(Ro), 
\label{eq:uzm2_low}
\end{eqnarray}
where FT stands for any term involving fluctuations not due to inertial waves. Consequently, if inertial waves are to influence the anisotropy of the mean flow, they can only do so through the last two terms of (\ref{eq:ozm}). Their part in this process is measured by the ratio of their contribution to inertial terms $|\langle \mathbf u^\prime\cdot\nabla \omega_z^\prime\rangle|$ and $|\langle \boldsymbol{\omega}^\prime\cdot\nabla u_z^\prime\rangle|$ (whose estimates are denoted $|\langle \mathbf u^\prime\cdot\nabla \omega_z^\prime\rangle|^e_{\rm IW}$ and $|\langle \boldsymbol{\omega}^\prime\cdot\nabla u_z^\prime\rangle|^e_{\rm IW}$) to the Coriolis term $|2\Omega\partial_z\langle u_z\rangle|$.
From figures \ref{fig10_fluc}c and d, the variations of these ratios with $Ro$ exhibit the same two regimes singled out in the analysis of (\ref{eq:uzm}). For $Ro\gtrsim2\times10^{-2}$, they are greater than unity and grow until they saturate at values between 10 and $10^2$. The saturation coincides with the regimes where inertial waves are superseded by random fluctuations (see figures \ref{fig4} and \ref{fig10_fluc}a), and marks the end of the rotation-dominated turbulent regime. 
Again, estimates are not representative of the behaviour of inertial waves in this range. Below the saturation, on the other hand, inertial waves dominate and (\ref{eq:ozm}) chiefly expresses a transfer of energy between them and the mean flow. For $Ro\lesssim 2\times10^{-2}$ and in the limit $Ro\rightarrow0$, the mean Coriolis term dominates in (\ref{eq:ozm}) and from (\ref{eq:uzm2_low}), it directly determines the flow anisotropy in the vertical plane according to
\begin{equation}
\left(\um\cdot \nabla \right)^2\ozm - \um\cdot \nabla\left( \langle\boldsymbol{\omega}\rangle\cdot \nabla \uzm \right)\simeq4\Omega^2\partial_{zz}^2 \Delta^{-1}\ozm-2\Omega\partial_z\um\cdot\nabla\uzm+ {\rm FT}.
\label{eq:ozm_lowro}
\end{equation}
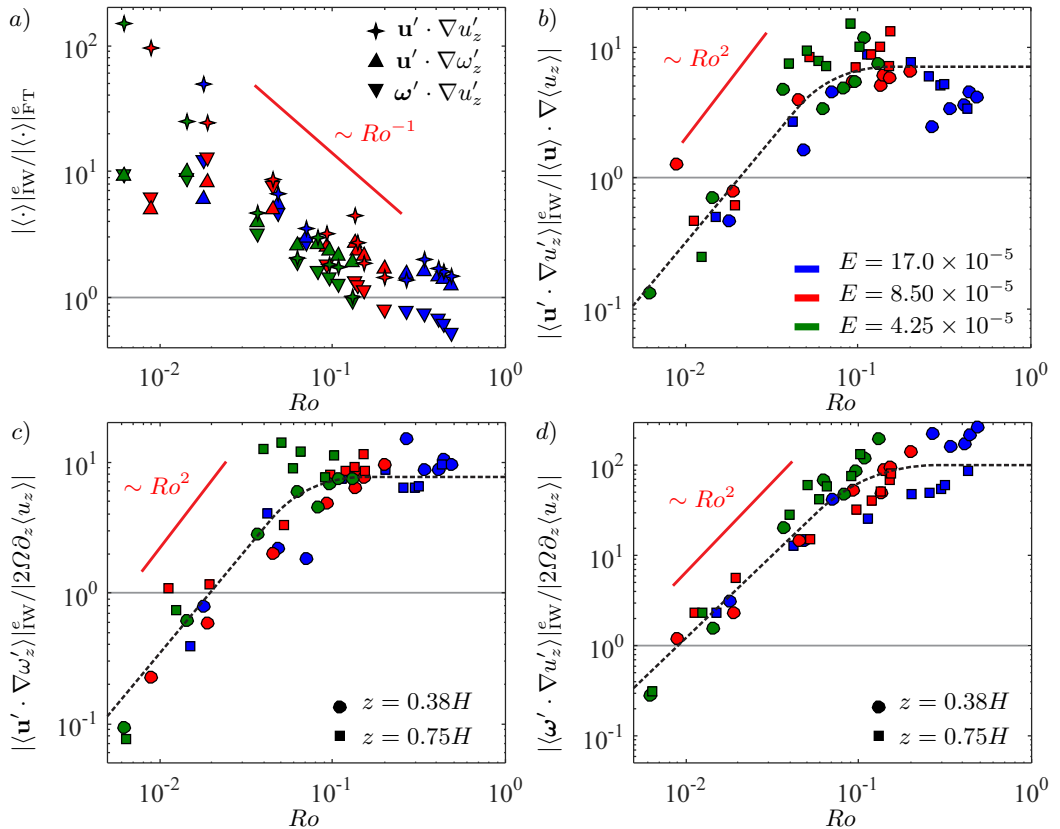
\begin{figure}
	\centerline{
		\includegraphics[scale = 0.95]{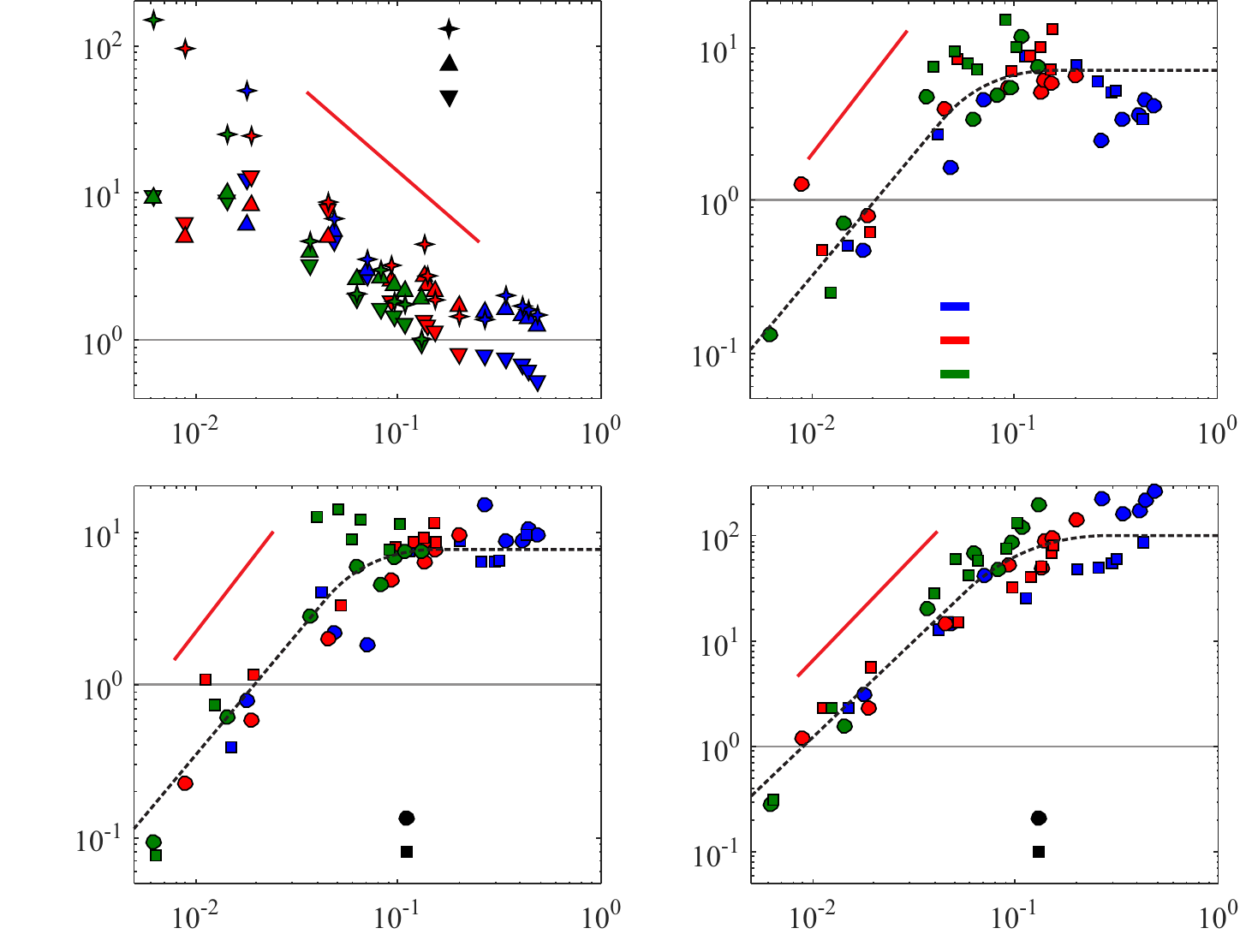}
		\put(-210,296){\makebox(0,0)[r]{\strut{}\footnotesize $\mathbf{u}^\prime\cdot\nabla u^\prime_z$ }}
		\put(-210,284){\makebox(0,0)[r]{\strut{}\footnotesize $\mathbf{u}^\prime\cdot\nabla\omega^\prime_z$ }}
		\put(-210,272){\makebox(0,0)[r]{\strut{}\footnotesize $\boldsymbol{\omega}^\prime\cdot\nabla u^\prime_z$ }}
		\put(-386,275){\rotatebox{-270}{\makebox(0,0)[r]{\strut{} \footnotesize $|\langle \cdot\rangle|^e_{\rm IW}/|\langle \cdot\rangle|^e_{\rm FT}$}}}%
		\put(-386,130){\rotatebox{-270}{\makebox(0,0)[r]{\strut{} \footnotesize $|\langle \mathbf{u}^\prime\cdot\nabla\omega^\prime_z \rangle|^e_{\rm IW}/|2\Omega\partial_z\langle u_z\rangle|$}}}%
		\put(-188,130){\rotatebox{-270}{\makebox(0,0)[r]{\strut{} \footnotesize $|\langle \boldsymbol{\omega}^\prime\cdot\nabla u^\prime_z \rangle|^e_{\rm IW} / |2\Omega\partial_z\langle u_z\rangle|$}}}%
		\put(-188,290){\rotatebox{-270}{\makebox(0,0)[r]{\strut{} \footnotesize $|\langle \mathbf{u}^\prime\cdot\nabla u^\prime_z\rangle|^e_{\rm IW}/ |\langle \mathbf{u}\rangle\cdot\nabla\langle u_z\rangle|$}}}%
		\put(-12,44){\makebox(0,0)[r]{\strut{}\footnotesize $z=0.38H$ }}
		\put(-12,31){\makebox(0,0)[r]{\strut{}\footnotesize $z=0.75H$ }}
		\put(-214,44){\makebox(0,0)[r]{\strut{}\footnotesize $z=0.38H$ }}
		\put(-214,31){\makebox(0,0)[r]{\strut{}\footnotesize $z=0.75H$ }}
		\put(-10,209){\makebox(0,0)[r]{\strut{}\footnotesize $E=17.0\times10^{-5}$ }}
		\put(-10,197){\makebox(0,0)[r]{\strut{}\footnotesize $E=8.50\times10^{-5}$ }}
		\put(-10,185){\makebox(0,0)[r]{\strut{}\footnotesize $E=4.25\times10^{-5}$ }}
		\put(-70,0){\makebox(0,0)[r]{\strut{} $Ro$ }}
		\put(-272,0){\makebox(0,0)[r]{\strut{} $Ro$ }}
		\put(-70,156){\makebox(0,0)[r]{\strut{} $Ro$ }}
		\put(-272,156){\makebox(0,0)[r]{\strut{} $Ro$ }}	
		\put(-115,120){\makebox(0,0)[r]{\strut{} \textcolor{red}{$\sim Ro^2$} }}
		\put(-320,125){\makebox(0,0)[r]{\strut{} \textcolor{red}{$\sim Ro^2$} }}
		\put(-117,285){\makebox(0,0)[r]{\strut{} \textcolor{red}{$\sim Ro^2$} }}
		\put(-235,257){\makebox(0,0)[r]{\strut{} \textcolor{red}{$\sim Ro^{-1}$} }}
		\put(-380,300){\makebox(0,0)[r]{\strut{} $a)$ }}
		\put(-182,300){\makebox(0,0)[r]{\strut{} $b)$ }}
		\put(-380,145){\makebox(0,0)[r]{\strut{} $c)$ }}
		\put(-182,145){\makebox(0,0)[r]{\strut{} $d)$ }}
	}
	\caption{a) Ratio between contributions by inertial waves fluctuations (IW) and random turbulent fluctuations (FT) to inertial terms in (\ref{eq:ozm}-\ref{eq:pmean}) at $z = 0.38H$. b) Ratio $|\langle \mathbf{u}^\prime\cdot\nabla u^\prime_z\rangle|^e_{\rm IW}/|\langle \mathbf{u}\rangle\cdot\nabla\langle u_z\rangle|$ at $z=0.38H$ (circles) and  $z=0.75H$ (squares). c) Ratio of $|\langle \mathbf{u}^\prime\cdot\nabla\omega^\prime_z\rangle|^e_{\rm IW}$ and d) $|\langle \boldsymbol{\omega}^\prime\cdot\nabla u^\prime_z\rangle|^e_{\rm IW}$ to Coriolis term $|2\Omega\partial_z\langle u_z\rangle|$. Black dashed lines show trend seen across the whole experimental parameter range. Same colour legend across all four graphs. Scalings are indicative only, they do not rely on any theoretical consideration at this stage.} 
	\label{fig10_fluc}
\end{figure}

\section{Conclusion and Discussion}
\label{sec:conclusion}
Eq.(\ref{eq:ozm_lowro}) establishes our main result: in the limit $Ro\rightarrow0$ inertial waves are not part of the mechanism driving the anisotropy of the average turbulent flow in background rotation. The actual mechanism involves a balance between advection of the average flow and the Coriolis force. From this point of view, it also differs from the diffusive process underpinning the formation of columns in Taylor's experiment. In both cases, however, anisotropy is materialised by a horizontally divergent flow either driven by inertia (turbulent flows) or by viscous friction (Taylor columns).\\ 
The physical mechanism  governing the anisotropy of the average flow along the rotation axis  is fully captured by (\ref{eq:ozm_lowro}), but more conveniently traced from the set of averaged equations (\ref{eq:ozm}-\ref{eq:pmean}), stripped from the contributions of inertial waves. As for the full flow field, the average flow $\langle \mathbf u\rangle$ is columnar in the limit $\Omega\rightarrow\infty$. 
Here, however, the departure to two-dimensionality for the average flow arises out of the inertial terms in (\ref{eq:ozm}), which presents itself in two forms: 
one is driven by inertia of the average flow itself (terms involving averages only), the other by inertia associated to fluctuations. Since we now know the contribution of inertial waves to these terms to be small, compared to the dominant Coriolis term in the limit $Ro\rightarrow0$, the divergent flow has to be predominantly driven by the other contributions to inertia: either from the average flow, or from random turbulent fluctuations. In turn, the vertical flow sustained by this mechanisms, drives an average vertical pressure gradient through (\ref{eq:uzm}), and subsequently, a $z-$dependence on the vertical vorticity component through (\ref{eq:pmean}), that defines its anisotropy.\\

This mechanism acts in two ways: the presence of inertia away from Ekman boundaries drives a departure to two-dimensionality through the secondary flows, with a mediating role played by the pressure. Conversely, if a localised horizontal force is applied to drive the flow (adding an inhomogeneous forcing term $\nabla\times\mathbf f$ in (\ref{eq:ozm})), then an average secondary flow with a vertical velocity component $\partial_z\langle u_z^f\rangle=-\frac1{2\Omega}\nabla\times\mathbf f$ is driven by the Coriolis force to be able to balance the rotational part of $\mathbf f$. Inertia then diverts the vertical flow into secondary flows, up to a distance $l_z$ from the location of the forcing, where the vertical flow driven by the forcing is exhausted. This point controls the vertical length scale (and anisotropy) of the average length scale set in motion by $\mathbf f$, as expressed by scaling (\ref{eq:lz_inertia}).\\
This mechanism explains how average anisotropy  can develop along the rotation axis without the involvement of inertial waves. It is closely related to the phenomenon of inertial mixing put forward by \cite{mcewan1976_nat}, but also generic in current-driven flows, whether this current involves mass or electric charge  (in MHD flows, \cite{p12_epl,pk2017_prf}). In magnetodhydrodynamic flows within the quasi-static approximation, Alfv\'en waves, which are the MHD counterpart of inertial waves, are altogether absent. Anisotropy develops through a diffusive process, in which the electric potential plays a very similar mediating role to that played by the pressure in rotating flows \citep{psm2000_jfm}. Similarly, electric current is driven along the magnetic field lines for the Lorentz force to balance the rotational part of inhomogeneous forces and the point where inertial forces have exhausted the current produced by them determines the length scale of the flow set in motion \citep{pk2017_prf}.\\

Finally, it is noteworthy that in the limit $Ro\rightarrow0$, $l^I_z(\l_\perp)/H\rightarrow\infty$ for all scales $l_\perp$, and the flow becomes quasi-two dimensional. Hence, our main result also implies that inertial waves are not responsible for the two-dimensionalisation of the average flow either. Thus the new mechanism argued for here may explain the formation of Taylor Columns in \emph{average} flows at high Reynolds numbers. We insist, however, that it does not concern Taylor columns found in the fluctuating part of the flow. In particular, flows with nil average flow, such as randomly forced turbulent flows in three-dimensional periodic domains may admit a basis of inertial modes. This implies that the formation of the fluctuating columns in these flows involves interactions between these waves, following either of the mechanisms proposed by \cite{smith99_pof}. By contrast, where a non-zero average exists, any basis must involve non-oscillatory modes (such as the leading order zonal flow in a sphere \citep{greenspan}). The experimental configuration studied in this paper falls in this category and it is precisely this steady component that is governed by the waveless mechanism we put forward. As such, this result does not exclude that in complex turbulent flows, the anisotropy of the average component may be driven by the waveless mechanism while at the same time anisotropic turbulent structures may be shaped by inertial waves.\\

The authors would like to acknowledge support from the Engineering and  Physical Sciences Research Council [grant number GR/N64519/01] for the manufacture of the rotating turntable facility and  B. Teaca for computational resources used in processing experimental data. AP acknowledges support from the Royal Society under the Wolfson Merit Award Scheme [Grant reference WM140032]. 

\bibliographystyle{jfm}
\bibliography{MASTER_LIB}

\end{document}